\documentclass[aps,showpacs,superscriptaddress,twocolumn,prb]{revtex4-1}
\usepackage{hyperref}
\usepackage{graphicx}
\usepackage{amsmath,amssymb,amsfonts}

\begin{document}

\title{Physical properties of single crystalline $R$Mg$_{2}$Cu$_{9}$ ($R$ = Y, Ce-Nd, Gd-Dy, Yb) and the search for in-plane magnetic anisotropy in hexagonal systems}

\author{Tai Kong}  
\affiliation{Department of Physics and Astronomy, Iowa State University, Ames, Iowa 50011, U.S.A.}
\affiliation{Ames Laboratory, U.S. DOE, Iowa State University, Ames, Iowa 50011, U.S.A.}
\author{William R. Meier}
\affiliation{Department of Physics and Astronomy, Iowa State University, Ames, Iowa 50011, U.S.A.}
\affiliation{Ames Laboratory, U.S. DOE, Iowa State University, Ames, Iowa 50011, U.S.A.}
\author{Qisheng Lin}
\affiliation{Ames Laboratory, U.S. DOE, Iowa State University, Ames, Iowa 50011, U.S.A.}
\author{Scott M. Saunders}
\affiliation{Department of Physics and Astronomy, Iowa State University, Ames, Iowa 50011, U.S.A.}
\affiliation{Ames Laboratory, U.S. DOE, Iowa State University, Ames, Iowa 50011, U.S.A.}
\author{Sergey L. Bud'ko}  
\affiliation{Department of Physics and Astronomy, Iowa State University, Ames, Iowa 50011, U.S.A.}
\affiliation{Ames Laboratory, U.S. DOE, Iowa State University, Ames, Iowa 50011, U.S.A.}
\author{Rebecca Flint}
\affiliation{Department of Physics and Astronomy, Iowa State University, Ames, Iowa 50011, U.S.A.}
\affiliation{Ames Laboratory, U.S. DOE, Iowa State University, Ames, Iowa 50011, U.S.A.}
\author{Paul C. Canfield}
\affiliation{Department of Physics and Astronomy, Iowa State University, Ames, Iowa 50011, U.S.A.}
\affiliation{Ames Laboratory, U.S. DOE, Iowa State University, Ames, Iowa 50011, U.S.A.}

\begin{abstract}

Single crystals of $R$Mg$_{2}$Cu$_{9}$ ($R$=Y, Ce-Nd, Gd-Dy, Yb) were grown using a high-temperature solution growth technique and were characterized by measurements of room-temperature x-ray diffraction, temperature-dependent specific heat and temperature-, field-dependent resistivity and anisotropic magnetization. YMg$_{2}$Cu$_{9}$ is a non-local-moment-bearing metal with an electronic specific heat coefficient, $\gamma \sim$ 15 mJ/mol K$^2$. Yb is divalent and basically non-moment bearing in YbMg$_{2}$Cu$_{9}$. Ce is trivalent in CeMg$_{2}$Cu$_{9}$ with two magnetic transitions being observed at 2.1 K and 1.5 K. PrMg$_{2}$Cu$_{9}$ does not exhibit any magnetic phase transition down to 0.5 K. The other members being studied ($R$=Nd, Gd-Dy) all exhibits antiferromagnetic transitions at low-temperatures ranging from 3.2 K for NdMg$_{2}$Cu$_{9}$ to 11.9 K for TbMg$_{2}$Cu$_{9}$. Whereas GdMg$_{2}$Cu$_{9}$ is isotropic in its paramagnetic state due to zero angular momentum ($L$=0), all the other local-moment-bearing members manifest an anisotropic, planar magnetization in their paramagnetic states. To further study this planar anisotropy, detailed angular-dependent magnetization was carried out on magnetically diluted (Y$_{0.99}$Tb$_{0.01}$)Mg$_{2}$Cu$_{9}$ and (Y$_{0.99}$Dy$_{0.01}$)Mg$_{2}$Cu$_{9}$. Despite the strong, planar magnetization anisotropy, the in-plane magnetic anisotropy is weak and field-dependent. A set of crystal electric field parameters are proposed to explain the observed magnetic anisotropy.

\end{abstract}

\maketitle

\section{Introduction}

Rare earth compounds are often studied for their magnetic properties when the rare earth ion is the only moment bearing element and when the rare earth fully occupies a single crystallographic site\cite{Taylor72, Szytula94, Canfield97, RNi, Myers99, Morosan04, Morosan05}. Magnetic anisotropies that are consistent with Heisenberg, Ising, and 4-state-clock models can be found originating from rare earth ions in the appropriate site symmetries\cite{Canfield97,Canfield97b,RNi,Myers99,Morosan04,Morosan05}. One of the interests is to look for in-plane magnetic anisotropy in a strongly planar system. Over the past decades, several studies had been carried out for systems with tetragonal symmetry, for example: HoNi$_{2}$B$_{2}$C\cite{Canfield97} and DyAgSb$_{2}$\cite{Myers99} where a 4-state clock model was realized. Several attempts had been made to find an analogy in a hexagonal symmetry\cite{Morosan04,Morosan05}. However, the local symmetries of rare earth ions in Ref. \onlinecite{Morosan04,Morosan05} are of orthorhombic m2m symmetry, even though the crystal structures have a hexagonal space group.

\begin{figure}[!h]
\centering
\includegraphics[scale=0.25]{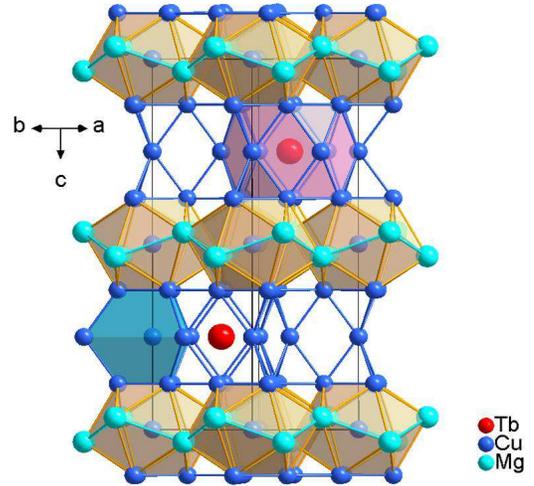}
\caption{(Color online) Unit cell of TbMg$_{2}$Cu$_{9}$. Elements are represented by solid spheres: Tb(red), Cu(blue) and Mg (cyan).}
\label{CS}
\end{figure}

The $R$Mg$_{2}$Cu$_{9}$ series of compounds were recently systematically synthesized and structurally identified\cite{Solokha06}. Their structure can be derived from the CeNi$_{3}$ type by replacing one of the two distinct rare earth sites with Mg atoms. As a consequence, there is only one rare earth site left in $R$Mg$_{2}$Cu$_{9}$ and it has a hexagonal site symmetry of $\bar{6}$m2. The environment of $R$ in $R$Mg$_2$Cu$_9$ is very similar to that in $R$Cu$_{5}$ and each layer that contains $R$ ions is separated from another by a layer of Cu-centered Cu$_6$Mg$_6$ icosahedra (see Fig.~\ref{CS}). $R$Mg$_{2}$Cu$_{9}$ is reported to exist for $R$ = Y, La-Nd, Sm-Ho, Yb. The reported lattice parameters follow a rough lanthanide contraction, except for possibly divalent Eu and Yb\cite{Solokha06}. 

Little has been characterized in terms of the physical properties for these compounds. Prior to the structural study, single crystals of CeMg$_{2}$Cu$_{9}$ were grown by melting and slow cooling of a stoichiometric composition\cite{Nakawaki02,Ito04}. It was reported to have an antiferromagnetic transition at 2.5 K. The magnetic transition temperature decreases with increasing pressure and seems to disappear at $\sim$2.5 GPa\cite{Nakawaki02,Ito04}. Polycrystalline EuMg$_{2}$Cu$_{9}$ seems to have a ferromagnetic transition at around 25 K\cite{Mauger10}. TbMg$_{2}$Cu$_{9}$ was studied as part of a search for hydrogen storage materials and was reported to order antiferromagnetically at around 10 K\cite{Pavlyuk11}. 

In $R$Mg$_2$Cu$_9$, since the rare earth is the only moment-bearing element and has one unique site, these compounds could potentially be good candidates to realize a 6-state-clock model. In this work, we present the results of structural measurements as well as temperature-dependent specific heat, temperature- and field-dependent electrical resistivity and temperature-, field- and angle-dependent magnetization on $R$Mg$_{2}$Cu$_{9}$ single crystals. Motivated by these results, in-plane magnetic anisotropy measurements on Y diluted TbMg$_{2}$Cu$_{9}$ and DyMg$_{2}$Cu$_{9}$ were made and will be discussed in the context of crystal electric field splitting.

\section{Crystal structure and experimental technique}

\begin{figure}[!h]
\begin{center}
\includegraphics[scale=0.36]{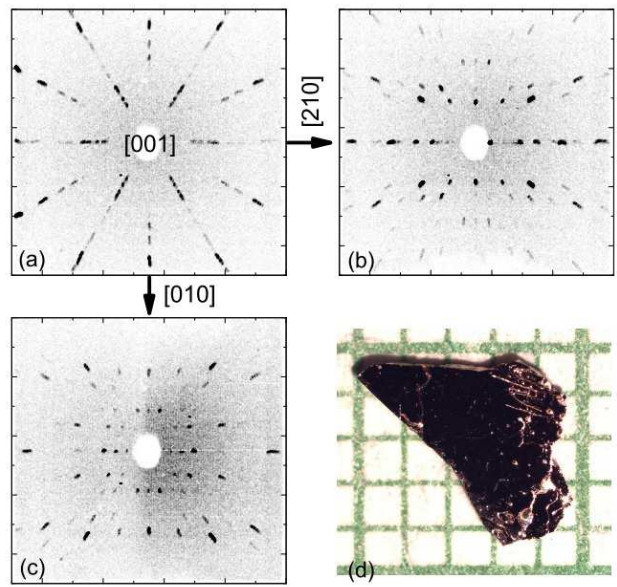}
\end{center}
\caption{(Color online) Laue pattern of $R$Mg$_{2}$Cu$_{9}$ along (a) [001] (b) [210] and (c) [010]. (d) Single crystal of PrMg$_{2}$Cu$_{9}$ on a millimeter grid paper. [001] is perpendicular to the facet shown.}
\label{Laue}
\end{figure}

Single crystals of $R$Mg$_{2}$Cu$_{9}$ were grown using a high-temperature solution growth method\cite{Canfield92}. Starting elements were held in a 3-cap tantalum crucible\cite{Canfield01} and sealed in a silica jacket under vacuum. Due to the complexity of the $R$-Mg-Cu ternary phase space, the starting stoichiometries vary. For $R$ = Ce-Nd, Tb, the starting elemental ratio was: $R$:Mg:Cu=2.5:20.4:77.1. For $R$=Y, Gd, Dy, Yb and Y with 1$\%$ Tb/Dy, the starting elemental ratio was : $R$:Mg:Cu=5:18:77. The ampoule assemblies were gradually heated up to 1180 $^{o}$C and decanted, after a 3-day slow cooling. For $R$ = Ce-Nd, Dy, the growths were decanted at 730 $^{o}$C; for $R$ = Y, Gd, the growths were decanted at 745 $^{o}$C; for $R$ = Tb, Yb, the growths were decanted at 760 $^{o}$C. Single crystals are pale-copper-metallic in color and plate-like with the crystallographic c-axis perpendicular to the plate. In Fig.~\ref{Laue}(d), a typical sample of PrMg$_{2}$Cu$_{9}$ is shown on a millimeter grid paper. A clear six-fold rotational symmetry can be seen from the Laue pattern when measuring along [001] [Fig.~\ref{Laue}(a)]. In-plane orientation in real space was also identified and corresponding Laue patterns are illustrated in Fig.~\ref{Laue}(b-c).

Crystallographic information was obtained by both single crystal x-ray diffraction and powder x-ray diffraction. Single crystal x-ray diffraction data were collected using a Bruker SMART APEX II CCD area-detector diffractometer\cite{SMART2003} equipped with a Mo K$_{\alpha}$ ($\lambda$ = 0.71073\AA) source. Integration of intensity data was performed by the SAINT-Plus program, absorption corrections\cite{Blessing1995} by SADABS , and least-squares refinements by SHELXL\cite{Sheldrick2000}, in the SMART software package. Powder x-ray diffraction data were collected using a Rigaku Miniflex II diffractometer at room temperature (Cu K$_{\alpha}$ radiation). Samples for powder x-ray diffraction were prepared by grinding single crystals into powders, after which powder was mounted on a single crystal Si, zero background sample holder with vacuum grease. Powder x-ray diffraction data were analyzed using the GSAS software\cite{Toby2001,GSAS}. Single crystal refinement data and atomic coordination information for TbMg$_{2}$Cu$_{9}$ are listed in Table~\ref{Tb129} and Table~\ref{AC}. A unit cell of TbMg$_{2}$Cu$_{9}$ is illustrated in Fig.~\ref{CS}.

\begin{table}[!h]
\caption{Single crystal crystallographic data for TbMg$_{2}$Cu$_{9}$ at room temperature.}
\begin{tabular}{p{5cm} p{3.8cm} }
\hline
\hline
Chemical formula & TbMg$_{2}$Cu$_{9}$\\
Formula weight (g/mol) & 779.40 \\
Space group & P6$_{3}$/mmc \\
Unit cell dimensions (\AA) & a = 5.0050(7) \\
 & c = 16.207(3)\\
Volume (\AA$^3$) & 351.59(12)\\
Z & 2\\
Density (g/cm$^{3}$) & 7.362 \\
Absorption coefficient (mm$^{-1}$) & 36.605\\
Reflections collected & 1571 [R(int)=0.0527]\\
Data/restraints/parameters & 213/0/8\\
Goodness-of-fit on F2 & 1.014\\
Final R indices [I $>$ 2sigma(I)] & R1=0.0385,wR2=0.0936\\
R indices (all data) & R1=0.0464, wR2=0.0983\\
Largest diff. Peak and hole (e/\AA$^{3}$) & 2.823 and -3.034\\
 
\hline
\end{tabular}
\label{Tb129}
\end{table}

\begin{table}[!h]
\caption{Atomic coordinates and equivalent isotropic displacement parameters for TbMg$_{2}$Cu$_{9}$ with full occupancy.}
\begin{tabular}{p{0.9cm} p{0.9cm} p{1.2cm} p{1.2cm} p{1.2cm} p{1.2cm} p{1.2cm}}
\hline
\hline
Atoms & Wyck. & Symm. & x & y & z & U$_{eq}$ (\AA$^{2}$)\\
Tb & 2d & -6m2 & 2/3 & 1/3 & 1/4 & 0.016(1)\\
Mg & 4f & 3m. & 2/3 & 1/3 & 0.4669(5) & 0.012(1)\\
Cu1 & 12k & .m. & 0.1682(2) & 2x & 0.3768(1) & 0.014(1)\\
Cu2 & 2c & -6m2 & 1/3 & 2/3 & 1/4 & 0.015(1)\\
Cu3 & 2b & -6m2 & 0 & 0 & 1/4 & 0.018(1)\\
Cu4 & 2a & -3m & 0 & 0 & 1/2 & 0.015(1)\\

\hline
\end{tabular}
\label{AC}
\end{table}

\begin{figure}[!ht]
\includegraphics[scale=0.35]{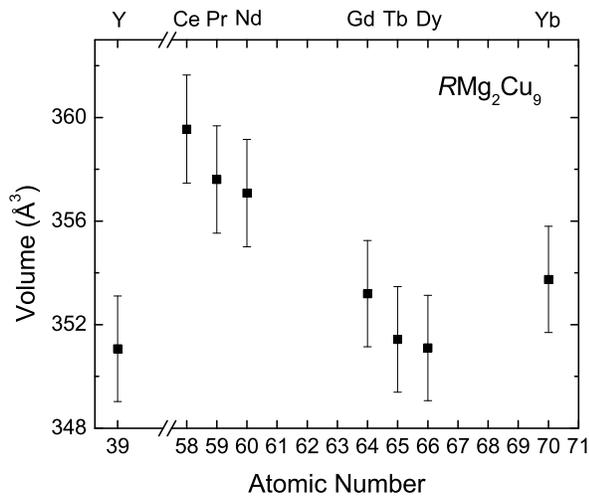}

\caption{Unit cell volume as a function of rare earth atomic number.}
\label{volume}
\end{figure}

Lattice parameters and unit cell volumes obtained from powder x-ray diffraction are listed in Table~\ref{lattice parameter} and unit cell volumes are plotted against the rare earth atomic number in Fig.~\ref{volume}. Generally, the lanthanide contraction is followed. YMg$_{2}$Cu$_{9}$ has a volume close to TbMg$_{2}$Cu$_{9}$ and DyMg$_{2}$Cu$_{9}$. The volume of YbMg$_{2}$Cu$_{9}$ is significantly larger than what would be expected from the lanthanide contraction for Yb$^{3+}$. This is consistent with the larger size of divalent Yb. Results from the current study agree with previously reported values\cite{Solokha06}.

\begin{table}[!h]
\caption{Lattice parameters and unit cell volumes of $R$Mg$_{2}$Cu$_{9}$ ($R$ = Y,Ce-Nd,Gd-Dy,Yb) obtained from powder x-ray diffraction. The uncertainty is about 0.2\% for lattice parameter values.}
\begin{center}
\begin{tabular}{|c|c|c|c|c|}
\hline
Compound & a (\AA) & c (\AA) & Volume (\AA$^3$) \\
\hline
YMg$_{2}$Cu$_{9}$& 5.00 & 16.19 & 351.1\\
\hline
CeMg$_{2}$Cu$_{9}$& 5.05 & 16.29 & 359.5\\
\hline
PrMg$_{2}$Cu$_{9}$& 5.04 & 16.26 & 357.6\\
\hline
NdMg$_{2}$Cu$_{9}$& 5.03 & 16.27 & 357.1\\
\hline
GdMg$_{2}$Cu$_{9}$& 5.02 & 16.21 & 353.2\\
\hline
TbMg$_{2}$Cu$_{9}$& 5.00 & 16.21 & 351.4\\
\hline
DyMg$_{2}$Cu$_{9}$& 5.00 & 16.20 & 351.1\\
\hline
YbMg$_{2}$Cu$_{9}$& 5.02 & 16.18 & 353.7\\
\hline

\end{tabular}
\end{center}
\label{lattice parameter}
\end{table}

Anisotropic dc magnetization up to 70 kOe was measured using a Quantum Design (QD) Magnetic Property Measurement System (MPMS). A QD Physical Property Measurement System (PPMS) was used to measure magnetization up to 140 kOe. Polycrystalline averaged magnetization was calculated from the equation: $\chi_{poly} = (\chi_c + 2\chi_{ab})/3$. $\chi_{poly}$ was also used to estimated the effective magnetic moment and to infer the magnetic ordering temperature from the peak temperature in d($\chi_{poly} T$)/d$T$\cite{Fisher62}. Angular-dependent dc magnetization was measured using a modified, QD, sample rotating platform with an angular resolution of 0.1$^{o}$. 

ac resistivity samples were prepared in a standard 4-probe geometry. Pt wires were attached to polished samples using Epotek-H20E silver epoxy. For the present study, electrical current was applied along [010] and magnetic field was applied along [210] as determined from the data in Fig.~\ref{Laue}(b-c). Both PPMS ($f$=17 Hz, $I$=1-3 mA) and Linear Research (LR), LR-700 ac resistance bridge ($f$=16 Hz, $I$=1-3 mA) were used to obtain resistivity data.

Specific heat was measured using a QD PPMS. A $^{3}$He option was used to obtain data down to 0.5 K. In order to estimate the magnetic specific heat, C$_{mag}$, associated with the local-moment-bearing members, the non-magnetic part of the specific heat, C$_{non-mag}$, was calculated based on the specific heat of YMg$_{2}$Cu$_{9}$ with the molar mass difference taken into account according to the Debye model\cite{Kittel}.

\section{Results}

\begin{figure*}[!ht]
\includegraphics[scale=0.65]{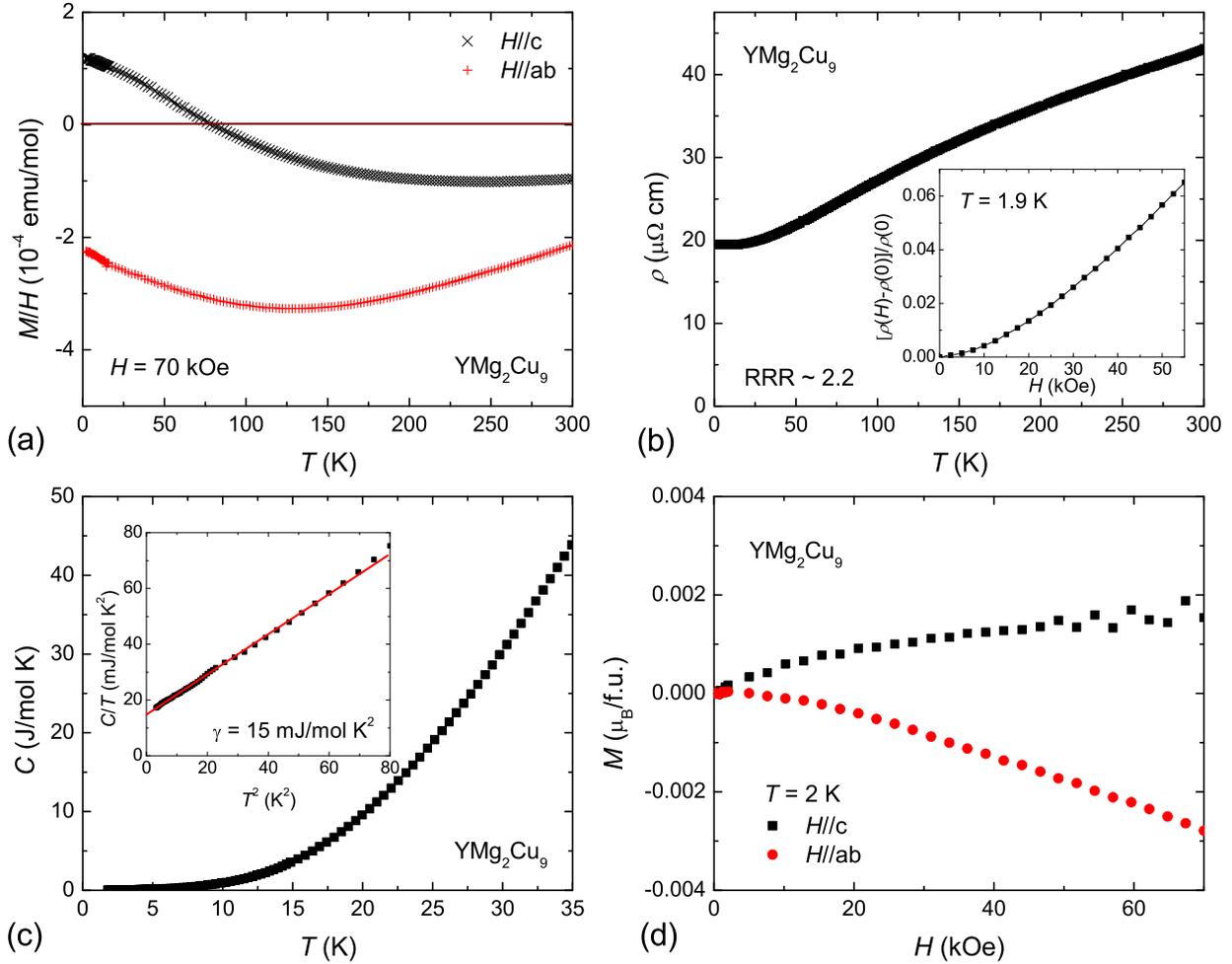}

\caption{(Color online) (a) Anisotropic temperature-dependent magnetization of YMg$_{2}$Cu$_{9}$ measured at 70 kOe. (b) Temperature-dependent resistivity. Inset: magnetoresistance measured at 1.9 K up to 55 kOe. (c) Temperature-dependent, zero-field specific heat (Inset: $C/T$ as a function of $T^{2}$. Red line shows the linear fit from base temperature to 9 K). (d) Magnetization isotherms measured at 2 K.}
\label{Y}
\end{figure*}

\subsection{YMg$_{2}$Cu$_{9}$}

Y does not have a 4f shell and bears no local-moment. Generally, a relatively temperature-independent magnetic susceptibility is expected due to Pauli paramagnetic, Landau diamagnetic, and core diamagnetic contributions. Details in Fermi surface may result in, albeit slight, magnetic anisotropies\cite{RNi,Kong14a}. The magnetic susceptibility of YMg$_{2}$Cu$_{9}$ is weakly temperature-dependent as shown in Fig.~\ref{Y}(a). The compound is diamagnetic at room temperature with $|\chi_c| < |\chi_{ab}|$. As temperature decreases, a broad minimum occurs at around 100 K, after which the magnetization increases. At 2 K, $\chi_c > 0$. This is consistent with field-dependent magnetization data measured at 2 K [Fig.~\ref{Y}(d)]. It is possible that some very low-level of magnetic impurities contribute to the low-temperature broad rise in magnetization as well as the non-linear, low-temperature $M$($H$) data.

The temperature-dependent specific heat data for YMg$_2$Cu$_9$ are shown in Fig.~\ref{Y}(c). From the linear fit of $C$/$T$ versus $T^2$, we estimated the Debye temperature to be around 320 K and the electronic specific heat, $\gamma$, to be around 15 mJ/mol-K$^2$ or $\sim$1 mJ/mole-atomic-K$^2$. 

The resistivity of YMg$_{2}$Cu$_{9}$ shows typical metallic behavior. The residual resistance ratio (RRR) is about 2.2. Magnetoresistance measured at 1.9 K roughly follows $H^{1.5}$ with an increase of 6$\%$ at 55 kOe.

\subsection{CeMg$_{2}$Cu$_{9}$}

\begin{figure*}[!t]
\includegraphics[scale=0.65]{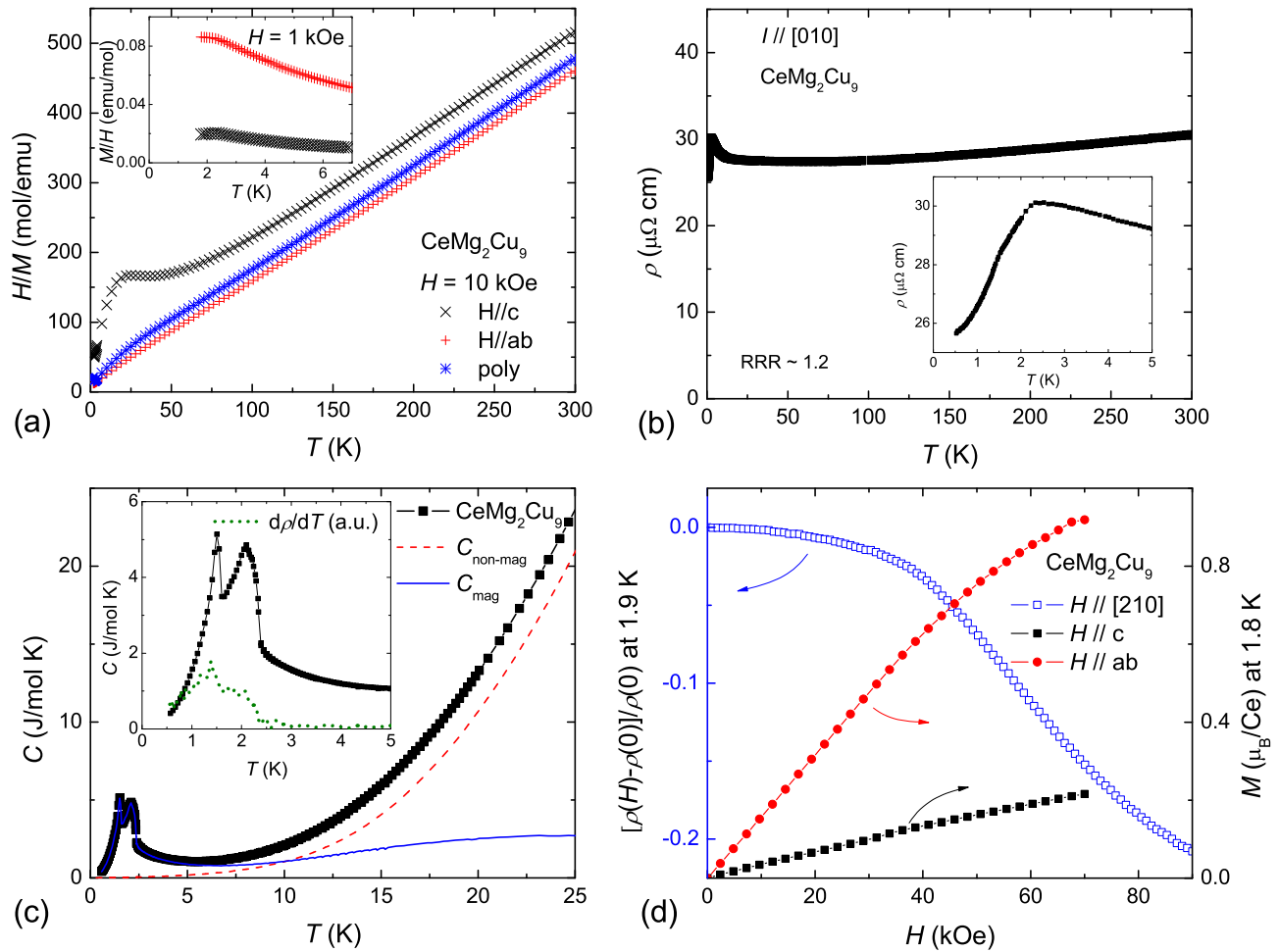}

\caption{(Color online) (a) Anisotropic inverse magnetic susceptibility of CeMg$_{2}$Cu$_{9}$ measured at 10 kOe (Inset: low-temperature magnetic susceptibility measured at 1 kOe). (b) Temperature-dependent resistivity (Inset: expanded view on the low temperature part of the resistivity). (c) Temperature-dependent, zero-field specific heat. Red dashed line and blue solid line represent C$_{non-mag}$ and C$_{mag}$ respectively. Inset: low temperature specific heat. Green dotted line represents d$\rho$/d$T$ in arbitrary units. (d) Magnetoresistance (blue) on the left and magnetization isotherms (black and red) on the right.}
\label{Ce}
\end{figure*}

The magnetic susceptibility of CeMg$_{2}$Cu$_{9}$ is anisotropic with $\chi_{ab} > \chi_c$. Fig.~\ref{Ce}(a) shows a very clear Curie-Weiss behavior (especially for $\chi_{poly}$) with estimated effective moment of Ce, $\mu_{eff}$ = 2.3 $\mu_B$, close to the theoretical value for Ce$^{3+}$ (2.5 $\mu_B$). The anisotropic Curie-Weiss temperatures are: $\Theta_c$ = -43 K, $\Theta_{ab}$ = -1 K and polycrystalline average $\Theta_{poly}$ = -12 K. At around 2 K, a change in $M(T)/H$ [inset of Fig.~\ref{Ce}(a)] suggests a possible antiferromagnetic transition. 

In addition to the reported magnetic transition at around 2 K\cite{Nakawaki02,Ito04}, one more phase transition at around 1.5 K was observed in the present study. The features that appear in resistivity data [Fig.~\ref{Ce}(b)] are consistent with the temperature-dependent specific heat data that are shown in Fig.~\ref{Ce}(c). The inset to Fig.~\ref{Ce}(c) shows C$_p$($T$), and d$\rho$/d$T$ data on an enlarged, low-temperature scale. Transitions at around 2.1 K and 1.5 K are apparent.

The electronic specific heat estimated above the transition temperature from 10 to 15 K is $\gamma \sim$ 58 mJ/mol-K$^{2}$, which is about 4 times higher than that for YMg$_{2}$Cu$_{9}$. It should be noted, though, that this value is smaller than previously reported values (115-160 mJ/mol-K$^{2}$)\cite{Nakawaki02,Ito04}. The discrepancy can be reduced by using the same temperature range of fitting; in between 8 and 10 K, the linear fit to $C/T$ versus $T^2$ gives a $\gamma$ value of $\sim$ 90 mJ/mol-K$^2$. However, in that temperature range, our data already show a certain degree of non-linearity. To this extent, for this compound, it is not clear if extracting a value for $\gamma$ is useful or constructive. In Fig.~\ref{Ce}(c), the red dashed line represents the non-magnetic part of the specific heat, C$_{non-mag}$, estimated from the specific heat of YMg$_2$Cu$_9$. Blue solid line represents the remaining, magnetic part of the specific heat, C$_{mag}$. Magnetic entropy estimated from C$_{mag}$ is close to Rln2 by the ordering temperature.

The temperature-dependent resistivity has a lower RRR ($\sim$1.2) than YMg$_{2}$Cu$_{9}$. It stays relatively constant down to 20 K, which could result from a distribution of local Kondo temperatures for a small number of the Ce sites affected by the disorder giving rise to the residual scattering\cite{Avila02}. The RRR of single crystals under study is lower than previously reported values. 

The field-dependent magnetization and resistivity at $\sim$1.8 K suggest a possible metamagnetic transition near 40 kOe. The metamagnetic transition is likely broadened because of the proximity of two phase transitions to the measurement temperature of 1.8 K. In the basal plane, the magnetization of 0.9 $\mu_B$/Ce at 70 kOe is nearly a half of the saturated moment of Ce$^{3+}$ (2.1 $\mu_B$). More metamagnetic transition could exist at higher applied magnetic fields as suggested by the magnetoresistance data.

\subsection{PrMg$_{2}$Cu$_{9}$}

\begin{figure*}[!ht]
\includegraphics[scale=0.65]{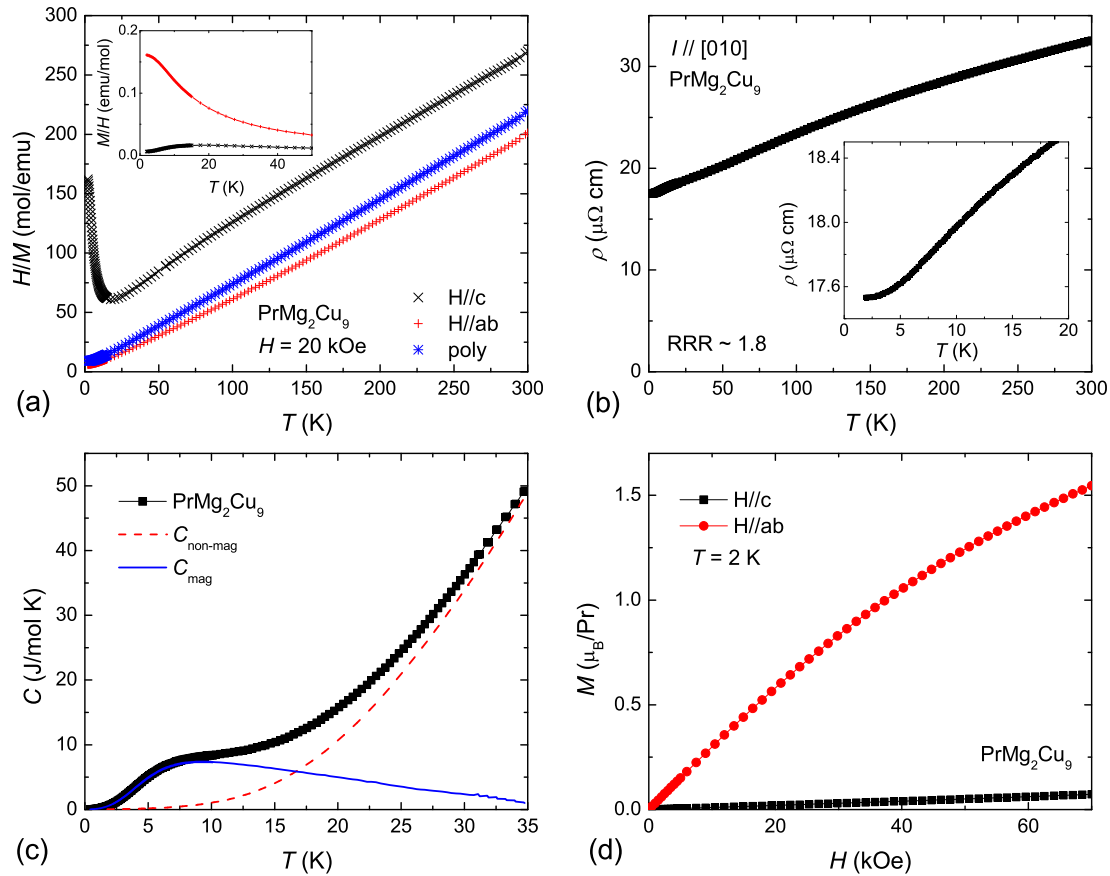}

\caption{(Color online) (a) Anisotropic inverse magnetic susceptibility of PrMg$_{2}$Cu$_{9}$ measured at 20 kOe. Inset shows the magnetic susceptibility at low temperature. (b) Temperature-dependent resistivity. Inset shows low temperature part of the resistivity. (c) Temperature-dependent, zero-field specific heat. Red dashed line and blue solid line represent C$_{non-mag}$ and C$_{mag}$ respectively. (d) Magnetization isotherms measured at 2 K.}
\label{Pr}
\end{figure*}

Data measured on PrMg$_{2}$Cu$_{9}$ single crystals are shown in Fig.~\ref{Pr}. The magnetization is anisotropic with $\chi_{ab} > \chi_c$. $\Theta_c$ =  -82 K, $\Theta_{ab}$ = 19 K and $\Theta_{poly}$ = 1 K. A linear fit of the polycrystalline averaged inverse magnetic susceptibility above 100 K yielded an effective moment of 3.3 $\mu_B$, close to the theoretical value of 3.6 $\mu_B$ for Pr$^{3+}$. As temperature decreases below 25 K, the magnetization seems to roll over to a non-magnetic ground state. No magnetic ordering was observed down to 2 K in magnetization. 

Specific heat of PrMg$_{2}$Cu$_{9}$ was measured down to 0.5 K and no phase transition was observed. At around 8 K, a broad dome in specific heat is consistent with a Schottky anomaly due to thermal population of excited CEF levels. The magnetic entropy increases to nearly Rln5 by 35 K. More discussions on the potential CEF level schemes will be presented in the next section.

The temperature-dependent resistivity of PrMg$_{2}$Cu$_{9}$ has a RRR of 1.8. The broad shoulder-like feature around 8 K coincides with the Schottky anomaly observed in the specific heat data. No signature of any ordering was observed down to the base temperature.

Fig.~\ref{Pr}(d) shows the field-dependent magnetization measured at 2 K. When the field is applied along c-axis, the magnetization slowly increase linearly with increasing field. For field along the ab-plane, the magnetization is much larger but the in-plane magnetization is still far from the saturation value (3.2 $\mu_B$) at 70 kOe.

\subsection{NdMg$_{2}$Cu$_{9}$}

\begin{figure*}[!ht]
\includegraphics[scale=0.65]{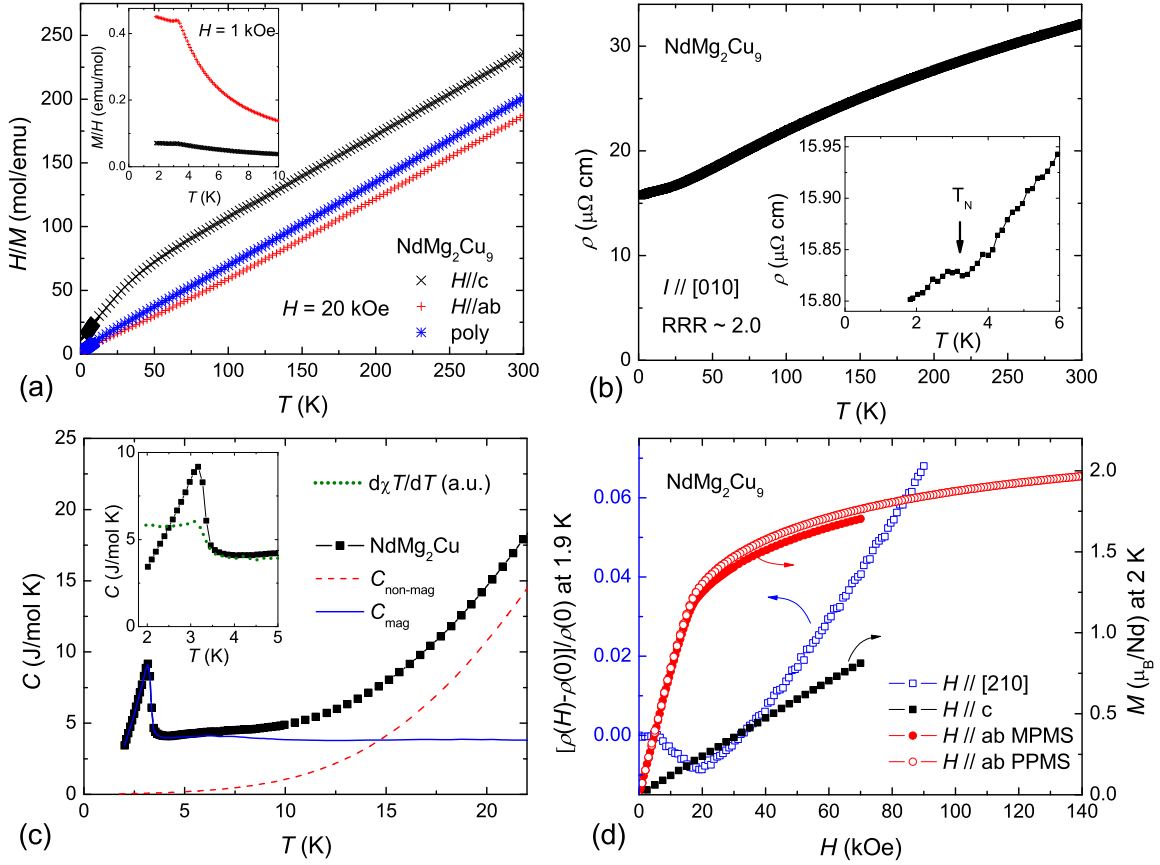}

\caption{(Color online) (a) Anisotropic inverse magnetic susceptibility of NdMg$_{2}$Cu$_{9}$ measured at 20 kOe (Inset: low-temperature magnetic susceptibility measured at 1 kOe). (b) Temperature-dependent resistivity (Inset: expanded view on the low temperature part of the resistivity. Arrow indicates the magnetic ordering temperature, T$_N$.). (c) Temperature-dependent, zero-field specific heat. Red dashed line and blue solid line represent C$_{non-mag}$ and C$_{mag}$ respectively. Inset shows an expanded view on the low-temperature specific heat. Green dotted lines represents d($\chi_{poly}T$)/d$T$ in arbitrary units. (d) Magnetoresistance (blue) on the left and magnetization isotherms (black and red) on the right.}
\label{Nd}
\end{figure*}

The magnetic anisotropy of NdMg$_{2}$Cu$_{9}$ in the paramagnetic state is similar to that observed in PrMg$_{2}$Cu$_{9}$ as shown in Fig.~\ref{Nd}(a). From the high-temperature, linear fit of inverse magnetic susceptibility, we obtained: $\Theta_{c}$ = -66 K, $\Theta_{ab}$ = 12 K and $\Theta_{poly}$ = -5 K. The calculated effective moment is 3.5 $\mu_B$ (theoretical value 3.6 $\mu_B$). A magnetic transition was observed at 3.2 K as featured by a kink in magnetization and a peak in d($\chi_{poly}T$)/d$T$ (Fig.~\ref{Nd}(c) inset). Below the magnetic ordering temperature, the magnetization becomes roughly temperature independent.

The temperature-dependent resistivity has a RRR of 2.0. As the temperature decrease down to the magnetic ordering temperature, the resistivity first increases slightly, suggesting a possible superzone gap opening due to magnetic ordering. The transition temperature inferred from magnetization and specific heat data is indicated in the inset of Fig.~\ref{Nd}(b) by a vertical arrow. The resistivity then continues decreasing at lower temperature. Clearer examples of a similar feature will be seen for TbMg$_2$Cu$_9$ and DyMg$_2$Cu$_9$ below.

Consistent with the magnetization data, the specific heat feature confirms a magnetic transition at 3.2 K. A small hump at around 7 K is most probably related to thermal population of excited CEF levels. Below the ordering temperature, an entropy of roughly Rln2 is removed. 

Both field-dependent magnetization and resistivity show a metamagnetic transition at around 20 kOe. The change of slope observed in magnetoresistance at 5 kOe with field applied along [210], however could not be well resolved in the magnetization data. Field-dependent, in-plane, magnetization was measured up to 140 kOe in order to search for more, high field, metamagnetic transitions, but none were observed. The magnetization reaches 2 $\mu_B$/Nd at 140 kOe, a value that is still not reaching the saturated value for Nd$^{3+}$ (3.3 $\mu_B$). It is possible that more metamangetic transitions may occur at higher applied field values. For field along the c-axis, the magnetization increases linearly with increasing field up to 70 kOe.

\subsection{GdMg$_{2}$Cu$_{9}$}

\begin{figure*}[!ht]
\includegraphics[scale=0.65]{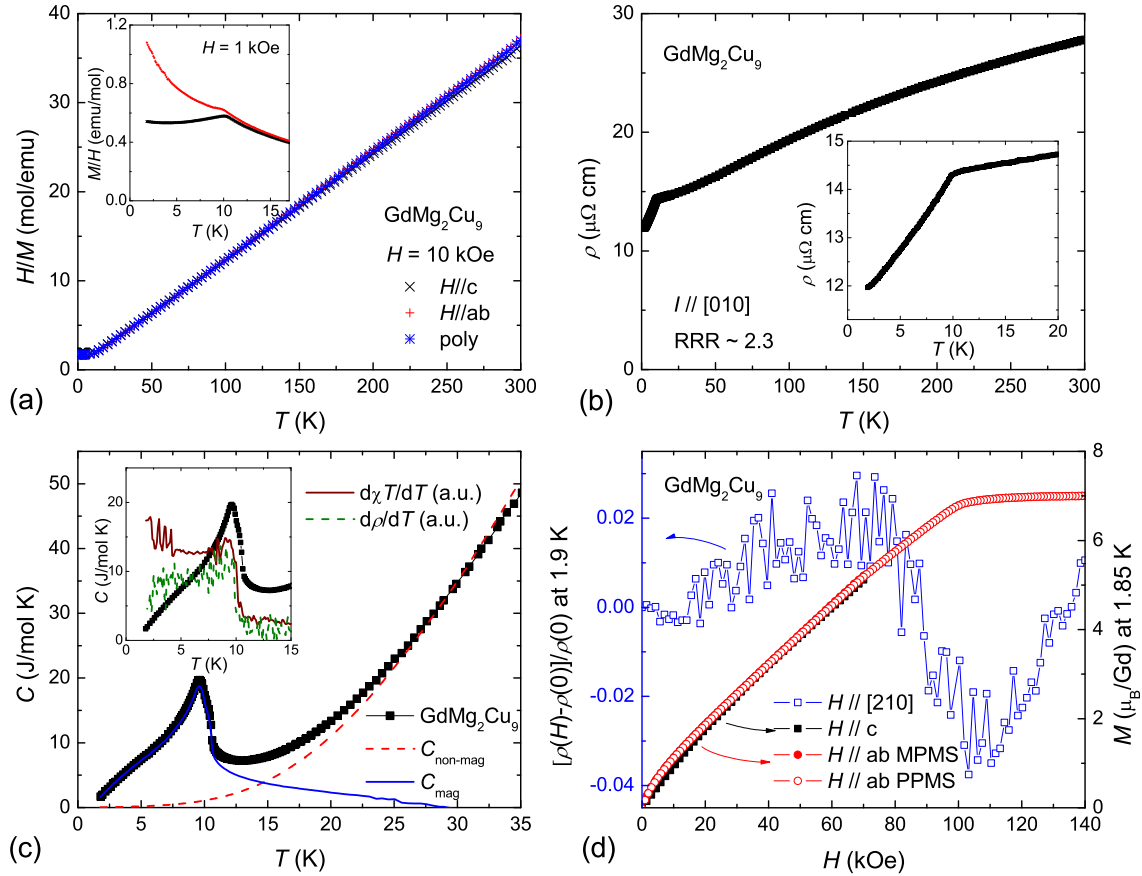}

\caption{(Color online) (a) Anisotropic inverse magnetic susceptibility of GdMg$_{2}$Cu$_{9}$ measured at 10 kOe (Inset: low-temperature magnetic susceptibility measured at 1 kOe. (b) Temperature-dependent resistivity (Inset: expanded view on the low temperature resistivity). (c) Temperature-dependent, zero-field specific heat. Red dashed line and blue solid line represent C$_{non-mag}$ and C$_{mag}$ respectively. Inset shows an expanded view on the low-temperature specific heat. Green dashed line (brown solid line) represents d$\rho$/d$T$ (d($\chi_{poly}T$)/d$T$) in arbitrary units. (d) Magnetoresistance (blue) on the left and magnetization isotherms (black and red) on the right.}
\label{Gd}
\end{figure*}

Because Gd$^{3+}$ has a half-filled 4f shell and thus zero angular moment, $L$=0, an essentially isotropic paramagnetic state is expected. Fig.~\ref{Gd}(a) shows just this for GdMg$_{2}$Cu$_{9}$, with $\Theta_{c}$=$\Theta_{ab}$=$\Theta_{poly}$ = -3 K. The effective moment is 8.1$\mu_B$, consistent with the expected value for Gd$^{3+}$ (7.9 $\mu_B$). Upon ordering near 10 K, the in-plane magnetic susceptibility keeps increasing and for field along the c-axis, magnetic susceptibility stays constant.  

The temperature-dependent resistivity [Fig.~\ref{Gd}(b)] shows a clear drop at the ordering temperature due to a loss of spin-disorder scattering. Unlike what was observed for NdMg$_{2}$Cu$_{9}$, no super-zone-gap-like feature was observed. The RRR of GdMg$_2$Cu$_9$ is around 2.3.

The temperature-dependent specific heat data [Fig.~\ref{Gd}(c)] does not show a clear $\lambda$-type anomaly and seem to suggest multiple transitions around 10 K. It first jumps at $\sim$10.5 K and then reaches a maximum at $\sim$9.6 K. Both d($\chi_{poly}T$)/d$T$ and d$\rho$/d$T$ show similar features. If only taking the peak positions in all three types of measurements, the magnetic transition temperature is at 9.7 K. The broad shoulder near around 5 K is common for Gd based compound and arises from a (2J+1)-fold degenerate multiplet\cite{Blanco91,Bouvier91, Kong14a}. Rln8 magnetic entropy is recovered by $\sim$ 17 K but since the non-magnetic part of the specific heat is not perfectly modelled by the YMg$_2$Cu$_9$ data, as evidenced by a crossing of C$_{non-mag}$ and total specific heat, the magnetic entropy inferred is only qualitative.

The field-dependent magnetization of GdMg$_2$Cu$_9$ [Fig.~\ref{Gd}(d)] is close to isotropic up to 70 kOe. The in-plane magnetization is only slightly larger than the out-of-plane magnetization below 20 kOe. A single metamagnetic transition was observed at around 100 kOe, above which the magnetic moment is saturated to 7 $\mu_B$/Gd, the same with theoretically predicted value. Magnetoresistance drops at the metamagnetic transition and increases with increasing field for both higher field and lower fields.

\subsection{TbMg$_{2}$Cu$_{9}$}

\begin{figure*}[!ht]
\includegraphics[scale=0.65]{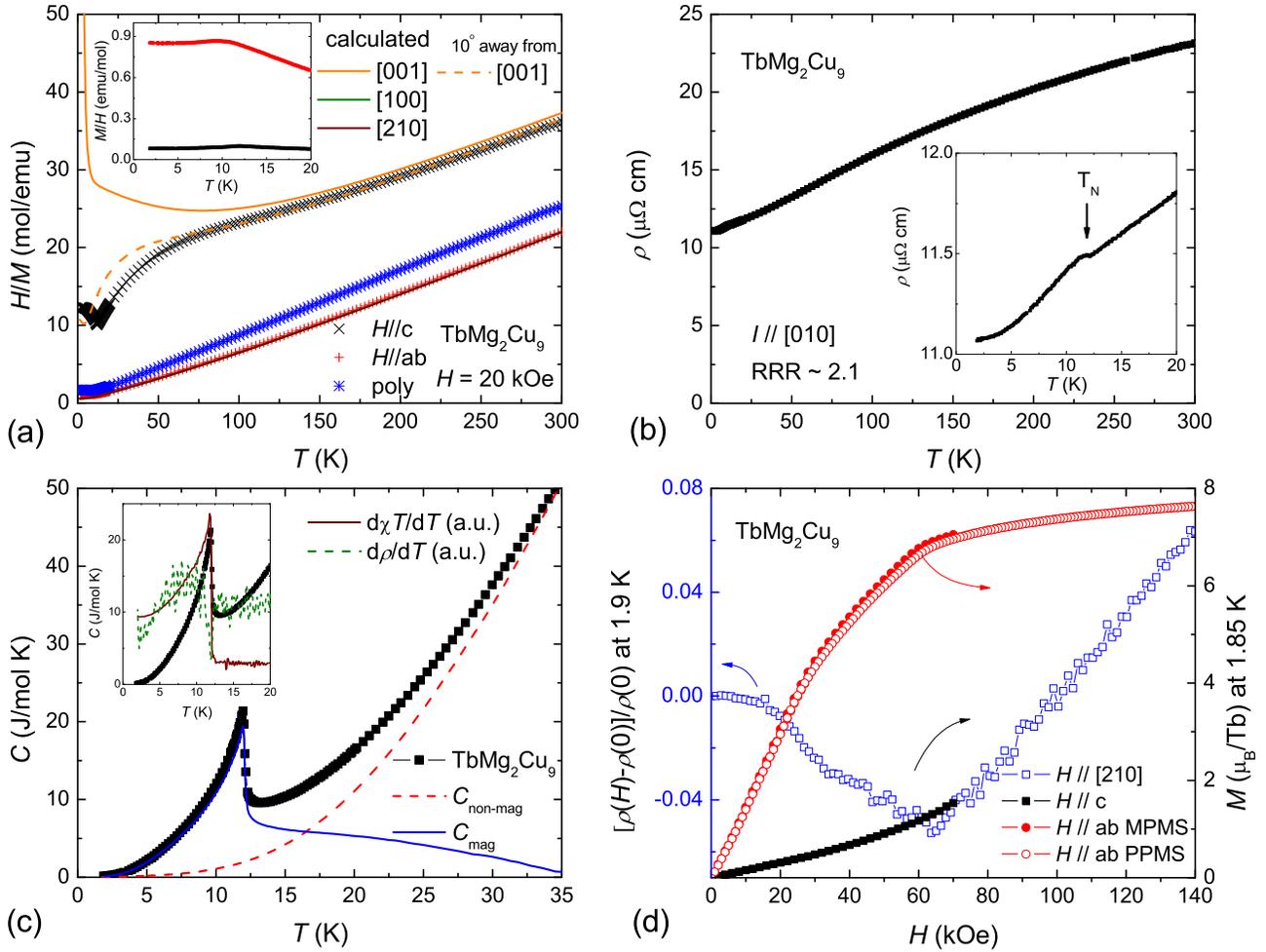}

\caption{(Color online) (a) Anisotropic inverse magnetic susceptibility of TbMg$_{2}$Cu$_{9}$ measured at 20 kOe (Inset: magnetic susceptibility at low temperature). Solid lines are calculated from proposed CEF levels (see text). (b) Temperature-dependent resistivity (Inset: low temperature part of the resistivity. Arrow indicates the magnetic ordering temperature, T$_N$.). (c) Temperature-dependent, zero-field specific heat. Red dashed line and blue solid line represent C$_{non-mag}$ and C$_{mag}$ respectively. Inset shows an expanded view on the low-temperature specific heat. Green dashed line (brown solid line) represents d$\rho$/d$T$ (d($\chi_{poly}T$)/d$T$) in arbitrary units. (d) Magnetoresistance (blue) on the left and magnetization isotherms (black and red) on the right.}
\label{Tb}
\end{figure*}

Data for TbMg$_{2}$Cu$_{9}$ are shown in Fig.~\ref{Tb}. The magnetization anisotropy of TbMg$_{2}$Cu$_{9}$ is strongly planar. A linear fit to the inverse magnetic susceptibility yields: $\Theta_c$ = -214 K, $\Theta_{ab}$ = 19 K and $\Theta_{poly}$ = 5 K. The inverse magnetic susceptibility of the polycrystalline averaged data remain linear down to a much lower temperature even though the CEF splitting leads to a much higher temperature anisotropic magnetic susceptibility\cite{Dunlap83}. The calculated effective moment is 9.8 $\mu_B$, close to expected value for Tb$^{3+}$ (9.7 $\mu_B$). 

The temperature-dependent resistivity of TbMg$_{2}$Cu$_{9}$ [Fig.~\ref{Tb}(b)] has a RRR of 2.1. Upon magnetic ordering, the resistivity shows behavior similar to NdMg$_{2}$Cu$_{9}$, suggesting the opening of a superzone gap. 

The specific heat, resistivity and magnetic susceptibility data all show consistent transition temperature values of T$_{N}$=11.9 K [Fig.~\ref{Tb}(c)]. The magnetic entropy was estimated to be close to Rln2 by the ordering temperature. Similar to GdMg$_2$Cu$_9$, magnetic entropy for TbMg$_2$Cu$_9$ upon ordering might be slightly different from that estimated here due to the inperfect nature of the YMg$_2$Cu$_9$ background subtraction, even after the mass correction for Tb instead of Y (see experimental methods). 

Metamagnetic transitions were observed in both electrical transport and magnetization measurements at $\sim$20 kOe and $\sim$60 kOe [Fig.~\ref{Tb}(d)]. For $H\perp$c, the magnetoresistance decreases below 60 kOe with a change in slope at 20 kOe. Above 60 kOe, the magnetoresistance increases linearly in field. The out-of-plane magnetization shows a weak up-curvature up to 70 kOe. The in-plane magnetization of TbMg$_{2}$Cu$_{9}$ is close to, but not yet reached the saturated moment of Tb$^{3+}$ (9 $\mu_B$) by 140 kOe.

\subsection{DyMg$_{2}$Cu$_{9}$}

\begin{figure*}[!ht]
\includegraphics[scale=0.65]{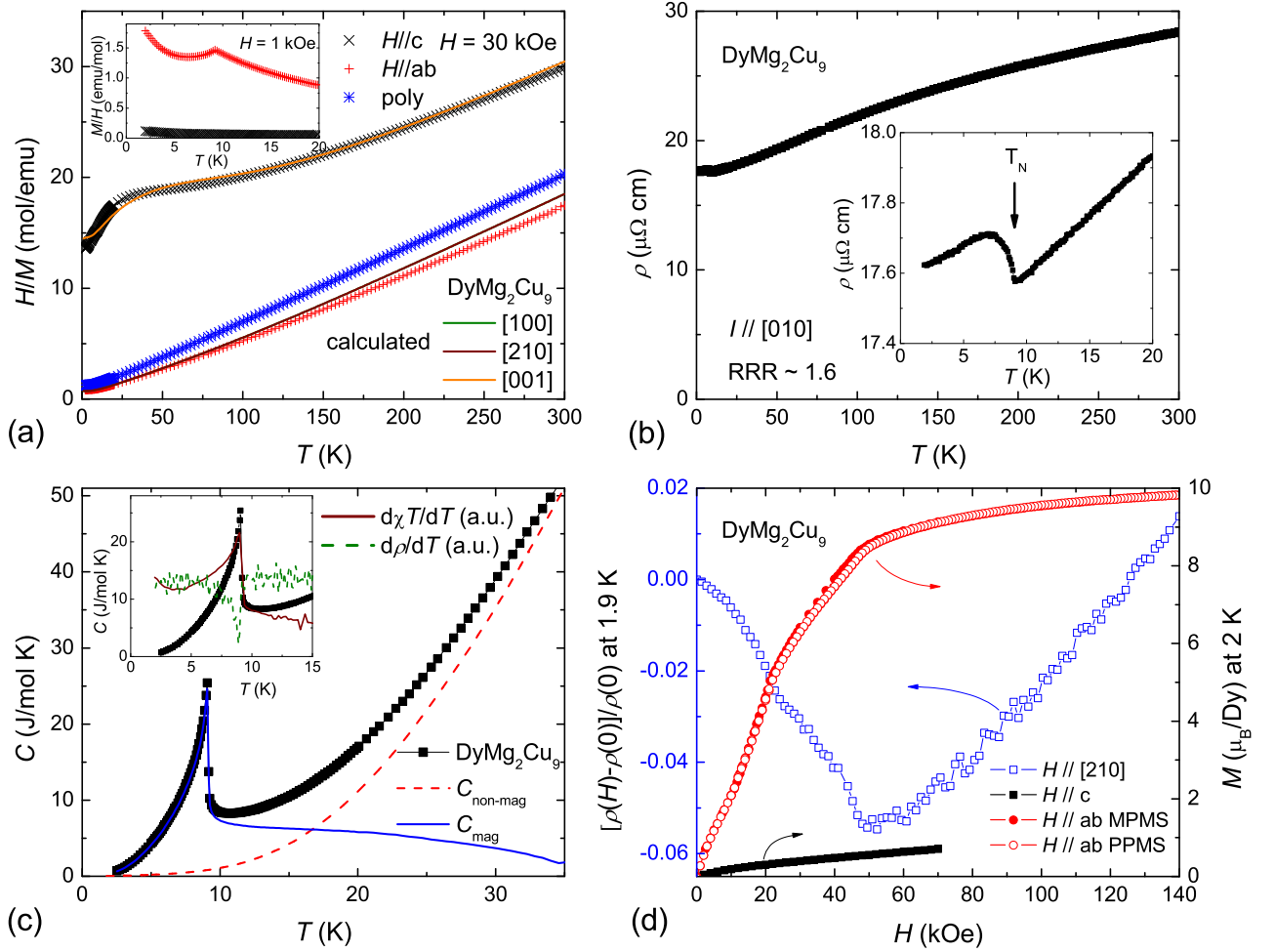}

\caption{(Color online) (a) Anisotropic inverse magnetic susceptibility of DyMg$_{2}$Cu$_{9}$ measured at 30 kOe (Inset: low-temperature magnetic susceptibility measured at 1 kOe). Solid lines are calculated from proposed CEF levels (see text). (b) Temperature-dependent resistivity (Inset: low temperature part of the resistivity. Arrow indicates the magnetic ordering temperature, T$_N$). (c) Temperature-dependent, zero-field specific heat. Red dashed line and blue solid line represent C$_{non-mag}$ and C$_{mag}$ respectively. Inset shows an expanded view of low-temperature specific heat. Green dashed line (brown solid line) represents d$\rho$/d$T$ (d($\chi_{poly}T$)/d$T$) in arbitrary units. (d) Magnetoresistance (blue) on the left and magnetization isotherms (black and red) on the right.}
\label{Dy}
\end{figure*}

The anisotropy of the temperature-dependent magnetization of DyMg$_{2}$Cu$_{9}$ [Fig.~\ref{Dy}(a)] is similar to that of TbMg$_{2}$Cu$_{9}$. $\chi_{ab}$ is much larger than $\chi_c$. At temperatures just above the magnetic ordering, $\chi_{ab}/\chi_c \sim $20. The Curie-Weiss temperatures extracted from inverse magnetic susceptibility are: $\Theta_c$ = -245 K, $\Theta_{ab}$ = 25 K and $\Theta_{poly}$ = -4 K. The effective moment is 10.9 $\mu_B$ (theoretical value: 10.6 $\mu_B$). Below $\sim$ 9 K, DyMg$_2$Cu$_9$ orders antiferromagnetically as suggested by the drop in magnetic susceptibility. 

As shown in Fig.~\ref{Dy}(b), the RRR of DyMg$_{2}$Cu$_{9}$ is about 1.6. A very clear increase of resistivity was observed at the magnetic transition temperature, similar to TbMg$_{2}$Cu$_{9}$.

Features for the magnetic transition in DyMg$_2$Cu$_9$ as seen from magnetic susceptibility and resistivity are consistent with $\lambda$-like anomaly in specific heat [Fig.~\ref{Dy}(c)]. The magnetic transition temperature T$_{N}$ is inferred to be 9.0 K. There is roughly Rln4 magnetic entropy removed below the ordering temperature.

There are two metamagnetic transitions observed at $\sim$20 kOe and $\sim$50 kOe [Fig.~\ref{Dy}(d)], above which the in-plane magnetization approaches the theoretical saturation value of 10 $\mu_B$. Overall, the field-dependent magnetization, specific heat and resistance are similar with what were observed in TbMg$_2$Cu$_9$. Magnetoresistance decreases with increasing field below 50 kOe and then increases linearly afterwards. The metamagnetic transition is marked by a change of slope. Magnetization along the c-axis increases monotonically up to 70 kOe.

\begin{figure}[!h]
\includegraphics[scale=0.31]{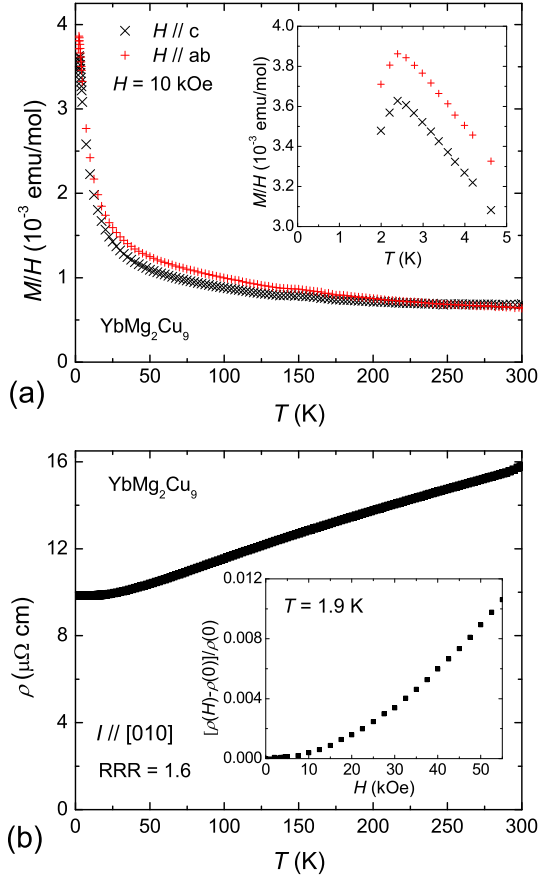}

\caption{(Color online) (a) Anisotropic magnetic susceptibility of YbMg$_{2}$Cu$_{9}$ measured at 10 kOe (inset: low-temperature magnetic susceptibility). (b) Temperature-dependent resistivity (inset: magnetoresistance measured at 1.9 K).}
\label{Yb}
\end{figure}

\subsection{YbMg$_{2}$Cu$_{9}$}

Given that the lattice parameters and unit cell volume of YbMg$_2$Cu$_9$ strongly deviate from the lanthanide contraction of $R^{3+}$ ions (Fig.~\ref{volume}), it is not surprising that the Yb ions appears to be Yb$^{2+}$. The temperature-dependent magnetization of YbMg$_{2}$Cu$_{9}$ is shown in Fig.~\ref{Yb}(a). The low-temperature Curie-tail and the cusp in magnetic susceptibility at around 2.5 K can be accounted for by about 0.5$\%$ molar contamination of Yb$_{2}$O$_{3}$ on or in the sample\cite{Li94}. The intrinsic magnetic susceptibility of YbMg$_2$Cu$_9$ can be inferred to be paramagnetic and temperature-independent with a magnitude of $\sim$ 5$\times$10$^{-4}$ emu/mol. This is comparable in magnitude to what was found for the non-magnetic YMg$_2$Cu$_9$. However, the difference in the core diamagnetism of Yb$^{2+}$ and Y$^{3+}$ is not sufficient to explain the exact change of the magnetic susceptibility\cite{Bain08}. The Fermi surfaces of the two are likely different due to an extra electron provided to the conduction band by Y$^{3+}$. 

Fig.~\ref{Yb}(b) shows the resistivity of YbMg$_2$Cu$_9$. It has a RRR value of 1.6. There is no indication of a phase transition down to 2 K. Magnetoresistance measured at 1.9 K increases by 1$\%$ 55 kOe.

\section{Trends across the $R$M\lowercase{g}$_2$C\lowercase{u}$_9$ series}

\begin{table}[!h]
\caption{Anisotropic Curie-Weiss temperatures ($\Theta_c$, $\Theta_{ab}$ and $\Theta_{poly}$), effective magnetic moment in paramagnetic state ($\mu_{eff}$) and magnetic transition temperatures ($T_{N}$) of $R$Mg$_{2}$Cu$_{9}$ ($R$ = Y, Ce-Nd, Gd-Dy, Yb). Magnetic transition temperatures are inferred from the peak temperatures of d($\chi_{poly} T$)/d$T$\cite{Fisher62}, d$\rho$/d$T$\cite{Fisher68} and specific heat.}
\begin{tabular}{p{1cm} p{1.2cm} p{1.5cm} p{1.5cm} p{1.5cm} p{1.5cm}}
\hline
\hline
$R$ & $\Theta_{c}$ (K)& $\Theta_{ab}$ (K)& $\Theta_{poly}$ (K)& $\mu_{eff}$ ($\mu_{B}$)& T$_{N}$ (K)\\
\hline
Y& -& -& -& -& -\\
Ce& -43 & -1 & -12 & 2.3 & 2.1, 1.5 \\
Pr& -82 & 19 & 1 & 3.3 & - \\
Nd& -66 & 12 & -5 & 3.5 & 3.2 \\
Gd& -3 & -3 & -3 & 8.1 & 9.7* \\
Tb& -214 & 19 & 5 & 9.8 & 11.9\\ 
Dy& -245 & 25 & -4 & 10.9 & 9.0 \\
Yb& - & - & - & - & -\\

\hline
\end{tabular}
*: There could be multiple transitions around this temperature up to 10.5 K.
\label{all data}
\end{table}

Anisotropic Curie-Weiss temperatures, effective moment in the paramagnetic state and the ordering temperatures for compounds under study are summarized in Table~\ref{all data}. Apart from isotropic GdMg$_2$Cu$_9$, the other local-moment-bearing compounds exhibit magnetic anisotropy with greater in-plane magnetization. The most extreme examples, TbMg$_2$Cu$_9$ and DyMg$_2$Cu$_9$, have $\chi_{ab}$ an order of magnitude larger than $\chi_c$ for $T \geq T_N$

\begin{figure}[!ht]
\includegraphics[scale=0.34]{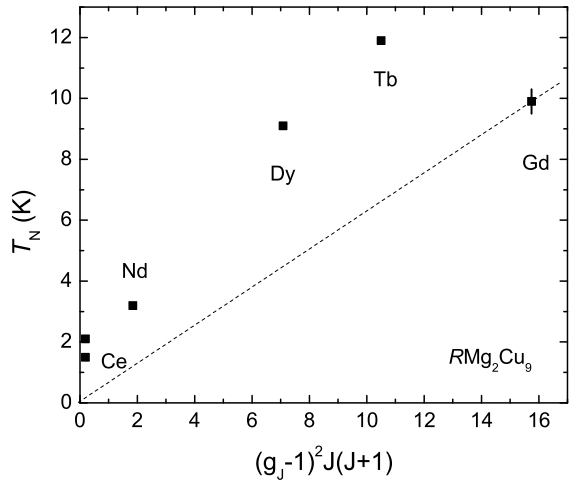}

\caption{Magnetic ordering temperatures of $R$Mg$_2$Cu$_9$ as a function of de Gennes factor $dG = (g_J-1)^2J(J+1)$. Potential multiple transitions for GdMg$_2$Cu$_9$ are represented by an error bar. Dotted line is a guide for the eyes.}
\label{summ}
\end{figure}

Ignoring the CEF effect and anisotropic exchange interaction, de Gennes argued that the Curie-Weiss temperatures and therefore, the magnetic ordering temperatures, in mean field theory, will be scaled with de Gennes factor\cite{dG62}: dG = $(g_J-1)^2J(J+1)$. The magnetic transition temperatures listed in Table.~\ref{all data} are plotted as a function of de Gennes factor in Fig.~\ref{summ}. As can be seen, such simple de Gennes scaling is not followed with TbMg$_2$Cu$_9$ and DyMg$_2$Cu$_9$ having higher transition temperatures than expected. In practice, both anisotropy in exchange interaction and CEF effect can, arguably, modify this scaling\cite{Hirst78,Noakes82}. In addition, since the magnetic anisotropy is mainly due to CEF effect, the strength of the exchange interaction that is responsible for low-temperature magnetic ordering may not be completely captured in the polycrystalline averaged Curie-Weiss temperatures (see Table~\ref{all data}). This may account for the inconsistency of $\Theta_{poly}$ values with de Gennes scaling. Experimentally, deviation from de Gennes scaling is not uncommon and has been observed in a variety of systems\cite{Dunlap84,Kong14a,Lin2013}. 

In the presence of the CEF effect, the ground state degeneracy will be lifted. In the case of PrMg$_2$Cu$_9$, both temperature-dependent magnetization and specific heat data are consistent with a singlet ground state. By fitting the low-temperature magnetization and specific heat data we can infer that PrMg$_2$Cu$_9$ likely has a singlet excited state at $\sim$12 K and a doublet at $\sim$ 25 K. As temperature drops below 20 K there is a gradual depopulation of the excited CEF levels that results in a rounded feature in temperature-dependent magnetization and a broad dome in $C_{mag}$. PrMg$_2$Cu$_9$, therefore, may be another example of a Pr-based intermetallic compound with a non-moment bearing, singlet CEF ground state\cite{Myers1999, Measson09}. 

The CEF splitting is also the dominating factor for the magnetic anisotropy that is observed in $R$Mg$_2$Cu$_9$. Quantitatively, the CEF Hamiltonian for the hexagonal rare earth site in this series can be written as\cite{Bauer09}:

\begin{equation}
H_{CEF} = B_2^0O_2^0 + B_4^0O_4^0 + B_6^0O_6^0 + B_6^6O_6^6
\label{Eq}
\end{equation}

\noindent
where B$_n^m$ are CEF parameters, O$_n^m$ are Steven operators\cite{Hutchings64, Bauer09}. In the point charge model, the CEF parameters can be expressed as B$_n^m$ = $A_n^m \langle r^n\rangle \theta_n$, where $\theta_2$ = $\alpha_j$; $\theta_4$ = $\beta_j$; $\theta_6$ = $\gamma_j$ are Steven's factors. $\langle r^n\rangle$ is the expectation value of the 4f radial function to the nth power. $A_n^m$ can often be viewed as a constant given the same crystalline neighboring environment. For uni-axial systems, B$_2^0$ is the leading term in determining the anisotropic Curie-Weiss temperatures, or in another word, being more planar or more axial\cite{Wang71}. Since $A_2^0$ does not change much from one rare earth to another in the same series of compounds and $\langle r^2 \rangle$ is always positive, the sign change of $\alpha_j$ will alter the axial/planar magnetic anisotropy. From theoretically calculated values for $\alpha_j$\cite{Hutchings64,Fulde85,Bauer09}, one can predict that the axial/planar magnetic anisotropy will be the same for $R$ = Ce-Nd, Tb-Ho trivalent ions. This is consistent with the data observed in $R$Mg$_2$Cu$_9$ series of compounds. In the following section, more detailed discussion on the CEF effect with respect to in-plane magnetic anisotropy will be presented.

\section{Angular dependent magnetization}

In rare earth compounds, the interplay between strong magnetic anisotropy and exchange interaction can often result in complex phase diagrams. For example, in strong axial systems, the Ising model with competing interactions was proposed to exhibit an infinite number of commensurate phases, also know as the devil's staircase\cite{Bak80}. Experimentally, many rare-earth-based systems, such as TbNi$_{2}$Ge$_{2}$\cite{RNi}, CeSb\cite{Rossat83,Wiener05}, TbNi$_{2}$Si$_{2}\cite{Blanco1991}$ have been studied as possible candidates. In the same manner, in strong planar systems, the 4-state-clock model was proposed in which moments in a tetragonal site symmetry are not only confined in-plane but also along a specific direction (an easy-axis)\cite{Canfield97,Kalatsky98}. Deviating from the easy-axis, the longitudinal magnetization decreases as a function of cos($\theta$) where $\theta$ is the angle between the direction of measurement and the nearest easy-axis. Such a model was motivated by and then used to understand complex phase diagrams and angular-dependent magnetization in tetragonal systems such as HoNi$_{2}$B$_{2}$C\cite{Canfield97} and DyAgSb$_{2}$\cite{Myers99}. Similarly, the complex phase diagrams of hexagonal compounds TbPtIn and TmAgGe have been interpreted in a modified 6-state-clock model based on three, crossed, in-plane Ising-like moments, caused by the orthorhombic site symmetry of the rare earth ions\cite{Morosan05}. A model system of a strongly planar, rare-earth-based compound with a hexagonal site symmetry has been, up to now, missing. In $R$Mg$_{2}$Cu$_{9}$, most of the investigated members in this study manifest promising features for such a study. There is a single rare earth site in a hexagonal site symmetry with a strong planar magnetization. Additionally, field-induced metamagnetic transitions were observed in all ordered members, even though not being very sharp compared to aforementioned 4-state-clock systems. Therefore, extremely planar members, DyMg$_{2}$Cu$_{9}$ and TbMg$_{2}$Cu$_{9}$, were examined in more detail with angular-dependent magnetization measurements.

\begin{figure}[!h]
\includegraphics[scale=0.27]{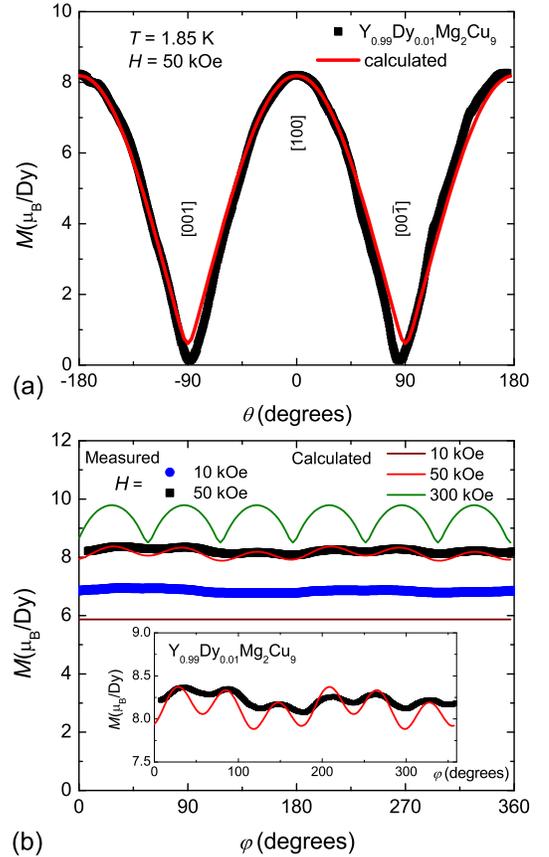}

\caption{Angular-dependent magnetization of Y$_{0.99}$Dy$_{0.01}$Mg$_{2}$Cu$_{9}$. (a) In-plane to out-of-plane magnetization measured at 1.85 K with 50 kOe. Red solid line represents results of CEF calculations. Field orientations are indicated by Miller indices. (b) In-plane magnetization measured at 1.85 K with 10 kOe, 50 kOe as a function of angle starting from [210]. Solid lines are calculated based on proposed single ion anisotropy of Dy due to CEF for $H$=10 kOe, 50 kOe and for comparison a hypothetical 300 kOe. Inset: expanded view of $H$=50 kOe data and results of CEF calculations.}
\label{ang_Dy}
\end{figure}

The confinement of the local-moments in plane is a critical requirement for the clock-type model. The CEF effect was considered as the primary contributor of such anisotropy for HoNi$_2$B$_2$C\cite{Canfield97} and DyAgSb$_2$\cite{Myers99}. Since CEF splitting is fundamentally a single ion effect, in order to better investigate the single ion magnetic anisotropy due to CEF splitting, 1$\%$ Dy or Tb was substituted into non-magnetic YMg$_2$Cu$_9$ in order to minimize the influence of magnetic interaction between local moments. In Fig.~\ref{ang_Dy}, both in-plane to out-of-plane, as well as purely in-plane, angular-dependent magnetization measurements are shown for Y$_{0.99}$Dy$_{0.01}$Mg$_2$Cu$_9$.

Fig.~\ref{ang_Dy}(a) shows a large axial-to-planar anisotropy. The in-plane magnetization is nearly two orders of magnitude larger than the out-of-plane magnetization. This is consistent with the magnetic anisotropy observed in pure DyMg$_2$Cu$_9$ (Fig.~\ref{Dy}). It also suggests that most of the magnetic anisotropy observed in the paramagnetic state comes from the single ion CEF effect. The in-plane anisotropy [Fig.~\ref{ang_Dy}(b)], on the other hand, is field-dependent and, even at $H$=50 kOe, only weakly angular-dependent. The magnetization measured at 10 kOe shows little indication of any 6-fold magnetic anisotropy. Under 50 kOe, the 6-fold modulation is only $\sim$ 3$\%$ of the total magnetization [inset in Fig.~\ref{ang_Dy}(b)]. A closer look at the data measured at both 10 kOe and 50 kOe reveals an additional 2-fold angular-dependence. This is likely due to an angular-dependent radial displacement of the sample from the centerline of the SQUID pick-up coil\cite{McElfresh96} associated with the sample mounting. Similar 2-fold modulation was observed for other systems when this rotating sample stage was utilized\cite{Fisher99}. 

Given that our experience has been that when a system has enough CEF splitting to manifest extreme planar anisotropy, it also manifests clear in-plane anisotropy in magnetic field\cite{Canfield97,Myers99,Morosan05}, these results require additional study. One possible, extrinsic cause of the coexistent of a strong planar and a weak in-plane magnetic anisotropy is a random twinning or crystalline domain formation such that the in-plane crystalline orientation is close to polycrystalline. This scenario was ruled out by conducting Laue measurements on different locations of the same crystal of sample at different depths (achieved by polishing). 

CEF splitting, without any extrinsic disorder, was then considered in order to explain the phenomena. In the presence of magnetic field, an additional Zeeman term, -$\vec{\mu}\cdot\vec{B}$, needs to be added into Equation~(\ref{Eq}). Magnetization under applied field can then be calculated based the derivative of free energy with field. Whereas B$_2^0$ determines axial/planar magnetic anisotropy, the mixture of different $J_z$ states is essential to the existence of in-plane magnetic anisotropy. In the current case, only applied magnetic field and $B_6^6O_6^6$ will mix different states. Worth noting, $B_6^6O_6^6$ by itself, or in a more general statement, pure CEF effects will not produce an in-plane magnetic anisotropy without magnetic field, be it externally applied or internal. More detailed examples will be illustrated and discussed below. 

The temperature-dependent magnetic susceptibility of pure DyMg$_2$Cu$_9$, rather than Y$_{0.99}$Dy$_{0.01}$Mg$_2$Cu$_9$, was used to compare with calculated values. This avoided problems caused by the weak temperature-dependence of YMg$_2$Cu$_{9}$'s magnetic susceptibility. And since the magnetic exchange interaction responsible for the low-temperature magnetic ordering of DyMg$_2$Cu$_9$ probably does not influence the magnetic anisotropy in its paramagnetic state by much (e.g. as can be seen in Table~\ref{all data}, $\Theta_{poly} \ll \Theta_{ab} \ll \Theta_{c}$), the magnetic susceptibility of pure DyMg$_2$Cu$_9$ is a good approximation to the single ion response. In Fig.~\ref{Dy}(a), solid lines show the calculated inverse magnetic susceptibility at 30 kOe with constraints of temperature-dependent magnetic entropy above the transition temperature, estimated from specific heat measurements. CEF parameters used are: B$_2^0$ = 1.99 K, B$_4^0$ = -1.00$\times$10$^{-4}$ K, B$_6^0$ = -1.70$\times$10$^{-5}$ K and B$_6^6$ = -7.50$\times$10$^{-4}$ K. The angular-dependent magnetization was then calculated based on this set of CEF parameters. In Fig.~\ref{ang_Dy}, solid lines show results of the calculated angular-dependent magnetization at various applied magnetic fields. For the in-plane to out-of-plane magnetization, the calculated values match with measured values quite well. 

The calculated in-plane anisotropy is very small at 10 kOe, consistent with what was observed even though the value of magnetization is smaller than the actual measured value. Taking into account a $|$cos$\theta|$-dependent radial displacement of the sample from the centerline of the pick-up coil during rotation with a maximum value of 2 mm, the calculated in-plane magnetic anisotropy at 50 kOe seems to qualitatively agree with experimentally observed results. The calculated magnetization at 300 kOe, as shown by green solid line, has a much clearer in-plane anisotropy. And the angular dependence is close to what would be expected if the moment can be simplified as a dipole with a preferred in-plane orientation (i.e. what would be called a 6-state-clock model). However, such a high magnetic field could not be accessed due to instrumental limitations.

\begin{figure}
\includegraphics[scale=0.85]{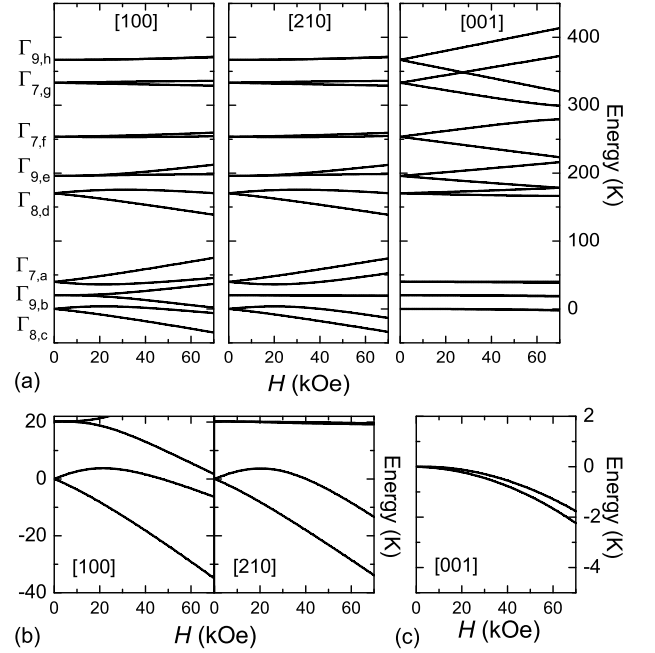}
\caption{Field-dependent CEF level energies for DyMg$_{2}$Cu$_{9}$ with $H$ along [100], [210] and [001]. The label on the left side was adapted from Ref.~\onlinecite{Segal1970}. An expanded view on the field-dependent CEF ground state with $H$ along [100] and [210] shown in (b), and along [001] shown in (c)}
\label{Dy_CEF}
\end{figure}

The origin of the anisotropic magnetization can be better understood by looking at the evolution of the CEF levels with applied field. In Fig.~\ref{Dy_CEF}, the field-dependent CEF levels for DyMg$_2$Cu$_9$ are plotted up to 70 kOe with applied field along three characteristic orientations for a hexagonal structure. In zero-field, the total CEF splitting is close to 350 K with the first excited and second excited states lying about 20 and 40 K above the ground state. Each CEF level of the Kramer's ion, Dy$^{3+}$, is a doublet. The labelling of each state follows the nomenclature used in Ref.~\onlinecite{Segal1970}. At low-temperatures, only the low-lying states contribute to the single ion magnetization. The ground state doublet, $\Gamma_{8,c}$, which is a mixture of $|\mp\frac{7}{2}>$ and $|\pm\frac{5}{2}>$, has a small moment along c-axis, as can be demonstrated by its splitting in applied field as shown in Fig.~\ref{Dy_CEF}(c). In addition, there is a mixing with $\Gamma_{8,d}$ that contributes to the c-axis magnetization, as manifested by the downward curvature of both of the Zeeman split halves of the ground state doublet. However, since all the levels with large c-axis magnetization values are high in energy and therefore not populated at any significant level at 1.85 K, the magnetization along c-axis is small. 

When field is applied in-plane, the CEF splitting only becomes markedly different above around 20 kOe where the first excited state $\Gamma_{9,b}$ evolves differently for the [100] and [210] directions. The mixing between $\Gamma_{9,b}$ and the ground state, $\Gamma_{8,c}$, plays an important role in the in-plane magnetic anisotropy. As a consequence, the variation of in-plane magnetization is small at 10 kOe. With increasing field, the difference in mixing among the low-lying states becomes more and more pronounced. This leads to a stronger in-plane anisotropy. The calculated in-plane magnetic anisotropy of the ground state ($\Gamma_{8,c}$) at 50 kOe is around 4$\%$ which is similar to the measured results.

A similar fit can be done for TbMg$_2$Cu$_9$ which gives B$_2^0$ = 3.38 K, B$_4^0$ = 4.12$\times$10$^{-4}$ K, B$_6^0$ = 1.88$\times$10$^{-5}$ K and B$_6^6$ = 8.48$\times$10$^{-4}$ K. In Fig.~\ref{Tb}(a), the modelled temperature-dependent magnetic susceptibility data at 20 kOe are shown together with the measured values. It qualitatively well matches experimentally obtained values above 150 K. Below 100 K, the calculated out-of-plane magnetization is smaller than measured data. This lower-temperature range can suffer from magnetic ordering and/or large, in-plane magnetization contribution from slightly misalignment of the sample. By a small, 10$^o$, misalignment from the c-axis as shown by the dashed line in Fig.~\ref{Tb}, the modelled magnetic susceptibility agrees much better with the experimentally measured data.

\begin{figure}
\includegraphics[scale=0.27]{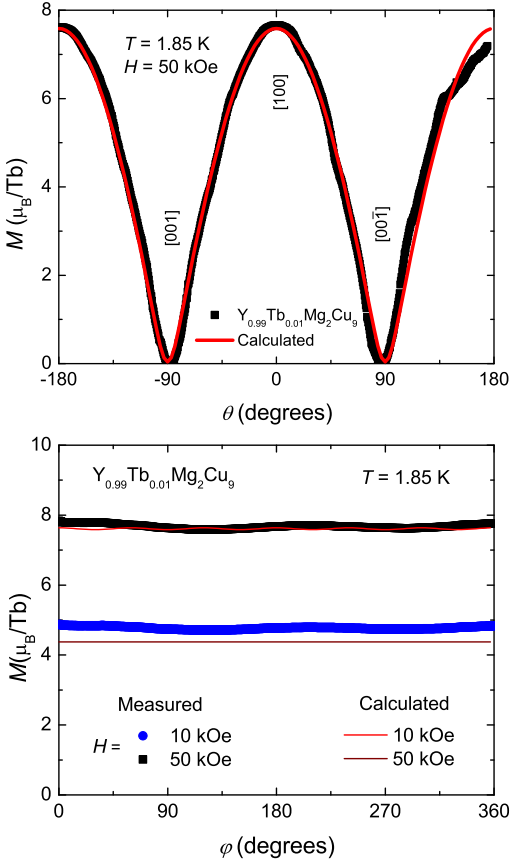}
\caption{Angular-dependent magnetization of Y$_{0.99}$Tb$_{0.01}$Mg$_{2}$Cu$_{9}$. (a) In-plane to out-of-plane magnetization measured at 1.85 K with 50 kOe. Red solid line represents results of CEF calculations. Field orientations are indicated by Miller indices. (b) In-plane magnetization measured at 1.85 K with 10 kOe, 50 kOe as a function of angle starting from [210]. Solid lines are calculated based on proposed single ion anisotropy of Tb due to CEF effect.}
\label{ang_Tb}
\end{figure}

\begin{figure}
\includegraphics[scale=0.85]{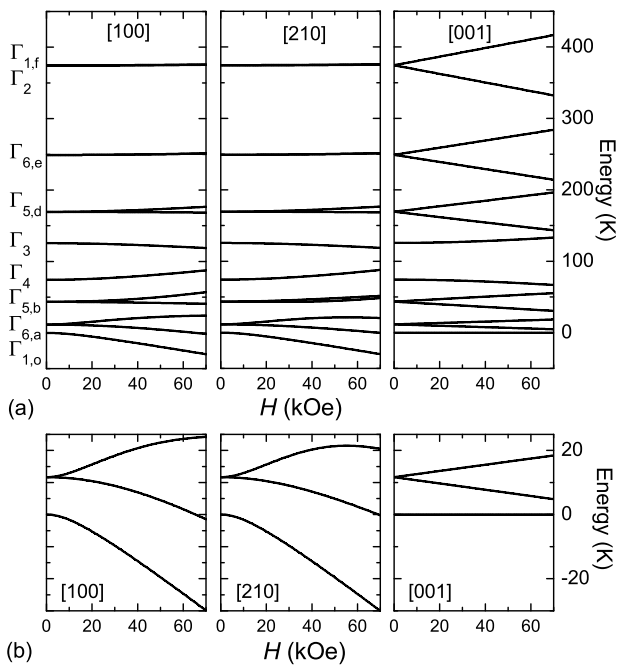}
\caption{Field-dependent CEF level energies for TbMg$_{2}$Cu$_{9}$ with $H$ along [100], [210] and [001]. The label on the left side was adapted from Ref.~\onlinecite{Segal1970}. (b) shows expanded views on the two lowest lying CEF levels with applied field in different directions.}
\label{Tb_CEF}
\end{figure}

Using the CEF parameters for Tb$^{3+}$, the angular-dependent magnetization can be calculated and compared to the experimental data shown in Fig.~\ref{ang_Tb}. As shown in Fig.~\ref{Tb_CEF}, the ground state of TbMg$_{2}$Cu$_{9}$, $\Gamma_{1,o}$, is a singlet. In addition, $\Gamma_{1,o}$ can only mix with $\Gamma_{1,f}$ and $\Gamma_{2}$ when field is applied along the c-axis. But as can be seen in Fig.~\ref{Tb_CEF}(a), these levels are very far above the ground state in energy, and therefore, provide only a small contribution to the magnetization. Even though the first and second excited states are moment-bearing doublets, they are not significantly populated at 1.85 K in 50 kOe (Fig.~\ref{Tb_CEF}(b)). With higher applied field, though, e.g. $\sim$90 kOe, $\Gamma_{6,a}$ will be closer to $\Gamma_{1,o}$ and become more populated at 1.85 K, which will result in an increased magnetization along c-axis. For the current case, at 1.85 K in 50 kOe, the magnetization along c-axis is nearly zero. That gives rise to the large in-plane to out-of-plane magnetic anisotropy. The measured in-plane to out-of-plane magnetization of Y$_{0.99}$Tb$_{0.01}$Mg$_2$Cu$_9$ resembles what was observed in Y$_{0.99}$Dy$_{0.01}$Mg$_2$Cu$_9$. The red solid line, representing the calculated value matches the experimental data very well. Note that the experimental magnetization value along [001] indeed goes toward zero for the systematically rotated sample, further suggesting that the low-temperature disagreement between data and CEF modelling in Fig.~\ref{Tb}(a) is due to slight misalignment.

In-plane angular-dependent magnetization measured at both 10 kOe and 50 kOe show little angular-dependence. This behavior is also well reproduced by the calculations and can be understood by considering the similar evolution of the low lying CEF levels with field in the [100] and [210] as shown in Fig.~\ref{Tb_CEF}. Even above 20 kOe, the majority of the ground state mixing is very similar between [100] and [210] that does not show a strong in-plane magnetic anisotropy. 

Based on theoretically calculated values for $\langle r^n\rangle$ and $\theta_n$ in the point charge model\cite{Hutchings64,Fulde85,Bauer09}, the CEF parameters for Tb$^{3+}$ can also be directly predicted from the values obtained for Dy$^{3+}$, which would be: B$_2^0$ = 3.33 K, B$_4^0$ = 2.27$\times$10$^{-4}$ K, B$_6^0$ = 2.09$\times$10$^{-5}$ K and B$_6^6$ = 9.21$\times$10$^{-4}$ K. These theoretically predicted parameters are close to the parameters that are directly obtained from experimental fitting shown above. Both sets of parameters give almost identical CEF level schemes as well as their field-dependences. The agreement between the point charge model prediction and the experimental fitted results in this case also partly validate our previous model used to understand the CEF effect in DyMg$_2$Cu$_9$. In general, angular-dependent magnetization observed here can be well modelled with CEF.

Comparing these results with HoNi$_2$B$_2$C and DyAgSb$_2$, where the 4-state-clock model is robust, TbMg$_2$Cu$_9$ and DyMg$_2$Cu$_9$ do not show comparable in-plane magnetic anisotropy under 50 kOe, even though the condition of a strong planar magnetization is satisfied. As described above, within the single ion picture, the CEF ground state will always have an isotropic, or XY-like, in-plane magnetization. It is only by mixing excited CEF levels in magnetic field that in-plane anisotropy can be realized. Of course, once the magnetic field becomes sufficiently strong, it will swamp the CEF splitting and remove any anisotropy, but that generally is at very large fields. Therefore there will be a "sweet spot", where the magnetic field can maximize the in-plane anisotropy. In TbMg$_2$Cu$_9$ and DyMg$_2$Cu$_9$, 50 kOe is very likely below that sweet spot. However, as shown in Fig.~\ref{ang_Dy}(b), the ideal max[cos($\theta$-n$\pi$/3] modulation of in-plane magnetization could potentially still be realized in DyMg$_2$Cu$_9$ for larger fields (300 kOe). 

On one hand, in this single ion situation, we can adjust the external applied magnetic field to find that sweet spot. On the other hand, in realizing a clock model in magnetically ordered compounds, the internal magnetic field due to exchange interaction, as a mean-field that originates from neighboring magnetic ions, can also induce the anisotropy. Essentially, the anisotropy of the moments ($\vec{S}_i$) depends on the effective magnetic field they feel ($\vec{H}_i$). Once the mean-field on each site, $\vec{H}_i = \sum_j J_{ij} \vec{S}_j(\vec{H}_j)$ becomes non-zero, below magnetic transition, they will develop an in-plane anisotropy, similar to $\vec{S}_i(\vec{H}_i)$ shown in the single-ion magnetization. This dependence leads to a slightly more complicated non-linear mean-field theory, where the energy from $H = \sum_{ij} J_{ij} \vec{S}_i(\vec{H}_i) \cdot \vec{S}_j (\vec{H_j})$ need to be minimized self-consistently with $\vec{H}_i = \sum_j J_{ij} \vec{S}_j(\vec{H}_j)$.

In searching for an in-plane state-clock model system, one needs a mixture of low-lying CEF states, which are well separated from higher-lying CEF states. In addition, in-plane magnetic anisotropy requires a very subtle balance between CEF effect, internal magnetic field and applied magnetic field. For example, decreasing the energy difference between low-lying states will increase the relative strength of a given applied magnetic field. In the case of DyMg$_2$Cu$_9$, by reducing the splitting between of the three lower-lying doublets ($\Gamma_{8,c}$, $\Gamma_{9,b}$, $\Gamma_{7,a}$) from 40 K to 10 K, a nearly ideal 6-state-clock state like that shown by green solid line in Fig.~\ref{ang_Dy}(b) can be realized at 50 kOe (as opposed to 300 kOe for the real compound). In the proposed CEF schemes for HoNi$_2$B$_2$C\cite{Cho96}, the 3 lowest-lying CEF levels in fact only have a span of $\sim$ 10 K and nearly 90 K away from higher CEF levels. This condition favors a clock-state-model at a moderate, reachable applied magnetic field as observed\cite{Canfield97}. However, angular-dependent magnetization data at different magnetic fields have not been measured on HoNi$_2$B$_2$C, nor on other 4-state-clock model systems\cite{Canfield97,Myers99}, which would be helpful to investigate the effect of this interplay between CEF and magnetic field on in-plane magnetic anisotropy.

\section{Conculsion}

Single crystals of $R$Mg$_2$Cu$_9$ ($R$= Y, Ce-Nd, Gd-Dy, Yb) have been synthesized using a high-temperature solution growth technique and characterized by magnetization, resistivity and specific heat measurements. YMg$_2$Cu$_9$ is non-magnetic. Ce is trivalent in CeMg$_2$Cu$_9$. It undergoes two magnetic transitions at 2.1 and 1.5 K respectively. PrMg$_2$Cu$_9$ does not order down to 0.5 K and appears to have a non-magnetic singlet ground state based on temperature-dependent magnetization and specific heat data. Yb is divalent, and therefore non-moment-bearing, in YbMg$_2$Cu$_9$. All the other local-moment-bearing members order antiferromagnetically at low-temperature. The ordering temperature of TbMg$_2$Cu$_9$ (11.9 K) is higher than that found in GdMg$_2$Cu$_9$ (9.7 K), indicating a deviation from de Gennes' scaling. Magnetic anisotropies were observed for $R$Mg$_2$Cu$_9$ ($R$ = Ce-Nd, Tb, Dy) with all of them showing a $\chi_{ab} > \chi_c$ in their paramagnetic states. Angular-dependent magnetization was studied in more detail for TbMg$_2$Cu$_9$ and DyMg$_2$Cu$_9$. Even though they have a strong planar magnetization, their in-plane magnetic anisotropy is small and field-dependent. This phenomena can be explained by single ion CEF effect where the laboratory magnetic field is not large enough to observe a clear clock-state given the CEF splitting. To observe an in-plane state-clock-model at low applied magnetic fields, the lower-lying CEF levels that can give rise to a large in-plane magnetization, as compared to out-of-plane magnetization, need to be relatively closely spaced in temperature and well separated from the higher-lying levels. Such a condition was met in the case of HoNi$_2$B$_2$C. However, a model system for a 6-clock-state is yet to be found.

\section*{Acknowledgement}

We would like to thank A. Kreyssig for useful discussions. Work done at Ames Laboratory was supported by US Department of Energy, Basic Energy Sciences, Division of Materials Sciences and Engineering under Contract NO. DE-AC02-07CH11358. W. R. Meier was funded by the Gordon and Betty Moore Foundation EPiQS Initiative through Grant GBMF4411. R. Flint was supported by the Ames Lab Royalty Fund and Iowa State Startup Funds.

\clearpage

\bibliographystyle{apsrev4-1}
%

\end{document}